\documentclass[journal=jpccck,manuscript=article,letterpaper,top=2cm,bottom=2cm,left=3cm,right=3cm,marginparwidth=1.75cm]{achemso}

\usepackage[english]{babel}
\usepackage{amsmath,amssymb}
\usepackage[export]{adjustbox}
\usepackage{graphicx}
\usepackage{gensymb}
\usepackage[colorlinks=true, allcolors=blue]{hyperref}
\usepackage[normalem]{ulem} 
\usepackage{xcolor}
\definecolor{lightblue}{RGB}{51,153,255}
\usepackage{ulem}
\usepackage[percent]{overpic}
\usepackage{float}
\usepackage{subcaption}
\usepackage[numbers,super,sort&compress]{natbib}

\title{Performance of Quadrupole Mass Filter with Tapered and Flared Geometry}

\author{Anushree Dutta}
\affiliation{School of Chemical Sciences, Indian Association for the Cultivation of Science, Kolkata-700032, India}

\author{Pintu Mandal}
\email{pintuphys@gmail.com}
\affiliation{Department of Physics, St. Paul's Cathedral Mission College, Kolkata-700009, India}

\author{Nabanita Deb}
\email{nabanita.deb@iacs.res.in}
\affiliation{School of Chemical Sciences, Indian Association for the Cultivation of Science, Kolkata-700032, India}

\begin{document}

\begin{tocentry}

\begin{figure}[H]
    \centering
    \includegraphics[width=0.8\linewidth]{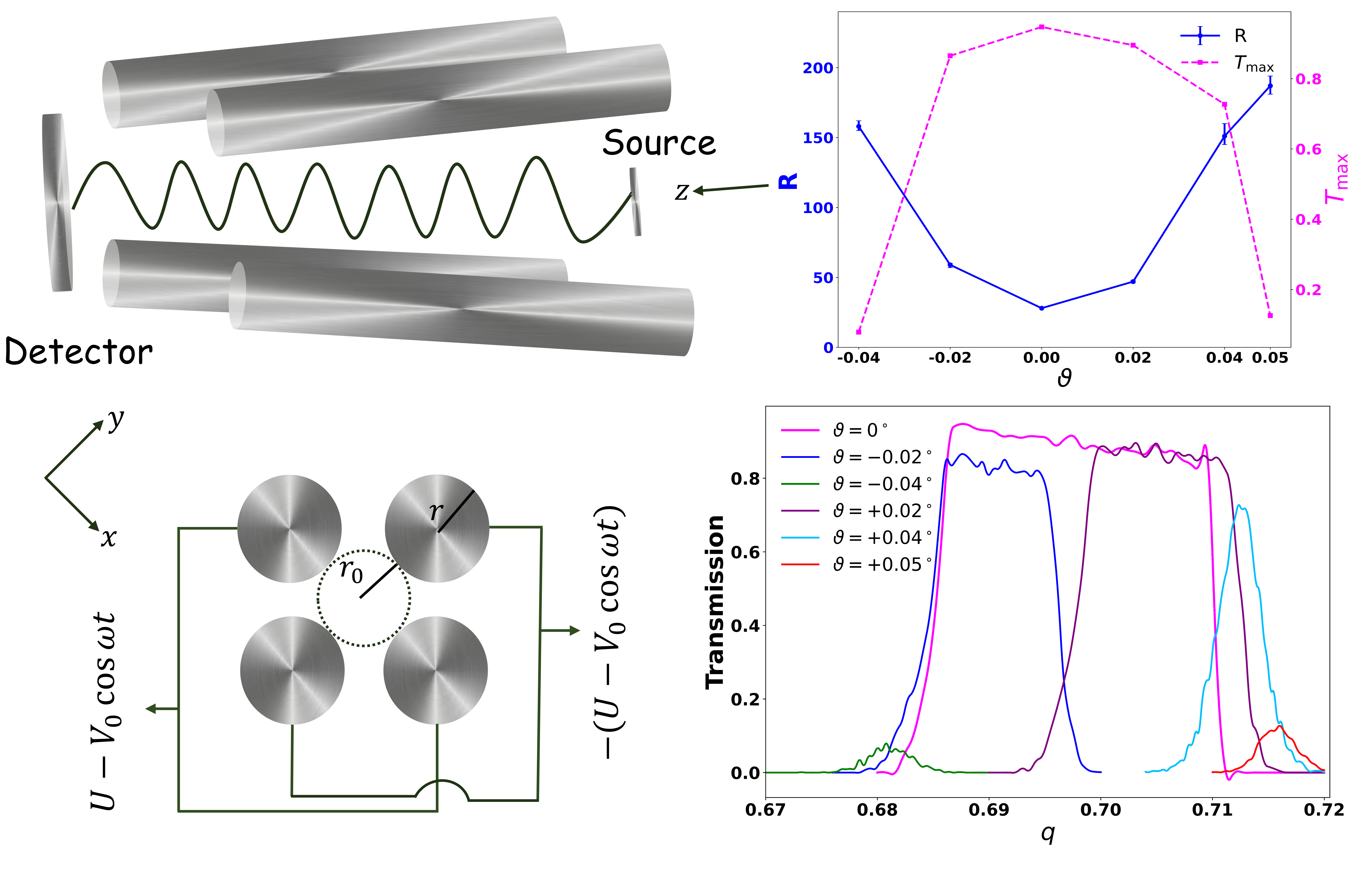}
    \caption*{}
    
    \label{fig:TOC}
\end{figure}
\end{tocentry}
\begin{abstract}
The performance of a quadrupole mass filter (QMF) is highly sensitive to deviations from ideal electrode geometry. In this work, we investigate the effect of small inward and outward tilting of cylindrical rods on the resolution and transmission characteristics of a QMF. Such geometric perturbations introduce an axial variation in the radial confinement potential, resulting in Mathieu parameters that vary along the ion trajectory. To examine this effect, the ion stability diagram is computed using a Runge–Kutta (RK45) method with axially-varying Mathieu parameters. The modified stability region exhibits shift and contraction depending on the magnitude and nature of rod inclination. The evolution of higher-order field components, particularly the dodecapole term, is analyzed along the axial direction. Ion trajectory simulations are performed using SIMION to evaluate the corresponding changes in QMF transmission characteristics in the first stability zone of operation. While simulations at fixed operating conditions indicate a transmission–resolution trade-off at small tilting angles leading to resolution enhancement, analysis at constant peak transmission reveals that even slight deviations from the parallel configuration lead to a degradation in resolution. These results highlight the critical role of minute geometric imperfections in QMF operation and provide insights into tolerance limits and design optimization for improved mass filter performance.

\end{abstract}

\textbf{keywords}: Quadrupole mass filter, tapered geometry, flared geometry, stability diagram, transmission characteristics, resolution

\section{Introduction} \label{sec1}

Quadrupole mass filters (QMFs) are fundamental components of modern mass spectrometry, offering a compact and efficient means of achieving mass-selective ion transmission \cite{paul1953neues, paul1955elektrische, dawson2013quadrupole, march1989quadrupole,  dayton1954measurement, denison1971operating, lee1971semi, reuben1994exact}. Their operation is governed by the stability of ion trajectories in combined RF and DC electric fields, described by the Mathieu equations, which define distinct stability regions in the ($a,q$) parameter space. The performance of a QMF, particularly its transmission efficiency, peak shape, and mass resolution, is determined by the precise positioning and sharpness of these stability boundaries.

Commercial QMFs typically employ circular rods arranged in a parallel configuration to approximate the ideal quadrupole potential required for radial confinement of ions with a selected mass-to-charge ratio. However, such electrode geometries inherently introduce higher-order multipole components, even in the absence of fabrication errors, which can influence device performance. In practice, additional deviations arise from mechanical imperfections, further distorting the electric field. A substantial body of work has therefore examined transverse asymmetries \cite{JANA2025117495, jana2025radial, ding2003quadrupole, taylor2008prediction,sysoev2022balance, mandal2024non, stafford1984recent}, including misalignment of rod pairs and variations in their effective radii, both of which alter the radial field distribution and consequently affect transmission characteristics and peak shapes.

In contrast, relatively few studies have addressed longitudinal asymmetry, wherein the effective field radius varies along the axial direction, leading to corresponding variations in the Mathieu parameters. Such conditions arise in configurations with tapered or flared rod geometries and have long been a concern in quadrupole design and fabrication. Axial variations in the radial electric field significantly influence ion dynamics and transmission behavior in QMFs. Even small geometric deviations can result in measurable changes in peak shape, transmission efficiency, and mass resolution~\cite{Dawson1988-1,Dawson1988-2}.

Dawson reported detailed calculations of QMF acceptance for finite-length systems, showing that axial variations in the Mathieu parameters—arising from non-parallel rod configurations or rapid scanning of operating parameters—can lead to peak shape irregularities~\cite{Dawson1988-1}. In a separate study, Dawson further demonstrated that even small deviations such as bent or bowed rod geometries can significantly affect ion acceptance, particularly for slow-moving ions under high-resolution conditions~\cite{Dawson1988-2}. These findings underscore the sensitivity of QMF performance to axial variations in the effective field geometry.

While earlier studies have largely focused on unintended geometric imperfections, controlled axial variations are increasingly being explored in emerging ion trap applications. Deliberately engineered geometries, such as tapered electrode configurations, enable spatial modulation of the trapping potential and have attracted significant interest. In particular, tapered or funnel-shaped traps have been proposed and realized in single-ion heat engines, where axial variation in confinement facilitates controlled energy exchange and thermodynamic cycles at the single-particle level \cite{Abah2012,Rossnagel2016,Dawkins2018,Levy2020}. In this context, Torrontegui \textit{et al.}~\cite{torrontegui2018transient} demonstrated a single-ion heat pump in a tapered trap, where axial motion dynamically modulates radial confinement to achieve controlled heat transfer. The use of shortcut-to-adiabaticity techniques, including transient non-confining potentials, enables faster operation while preserving stability. Beyond heat engines, tailored electrode geometries are also relevant for precise control of ion dynamics and trapping efficiency in advanced ion-based devices, underscoring the importance of understanding how axial variations in the effective field radius influence ion dynamics and stability~\cite{Deng2025}.

In this work, the influence of tapered and flared geometries on the performance of a QMF is systematically investigated within the first stability region. The axial variation of the quadrupole field is modeled by incorporating the position-dependent field radius, leading to a corresponding variation in the effective Mathieu parameters. To the best of our knowledge, stability diagrams for such progressively varying field radii have not been systematically examined, and this problem is addressed in the present work by computing the stability diagram under the assumption of constant longitudinal ion energy. Ion transmission is first analyzed under fixed operating conditions to elucidate the effect of longitudinal asymmetry on transmission efficiency, peak evolution, and resolving power. A complementary analysis at constant peak transmission further shows that even small deviations from the parallel rod configuration lead to a reduction in mass resolution. These results provide insight into the transmission–resolution trade-off in non-ideal geometries and offer practical guidance for the design and optimization of QMFs with controlled axial variations.

\begin{figure}[H]
    \centering
    \includegraphics[width=0.5\linewidth]{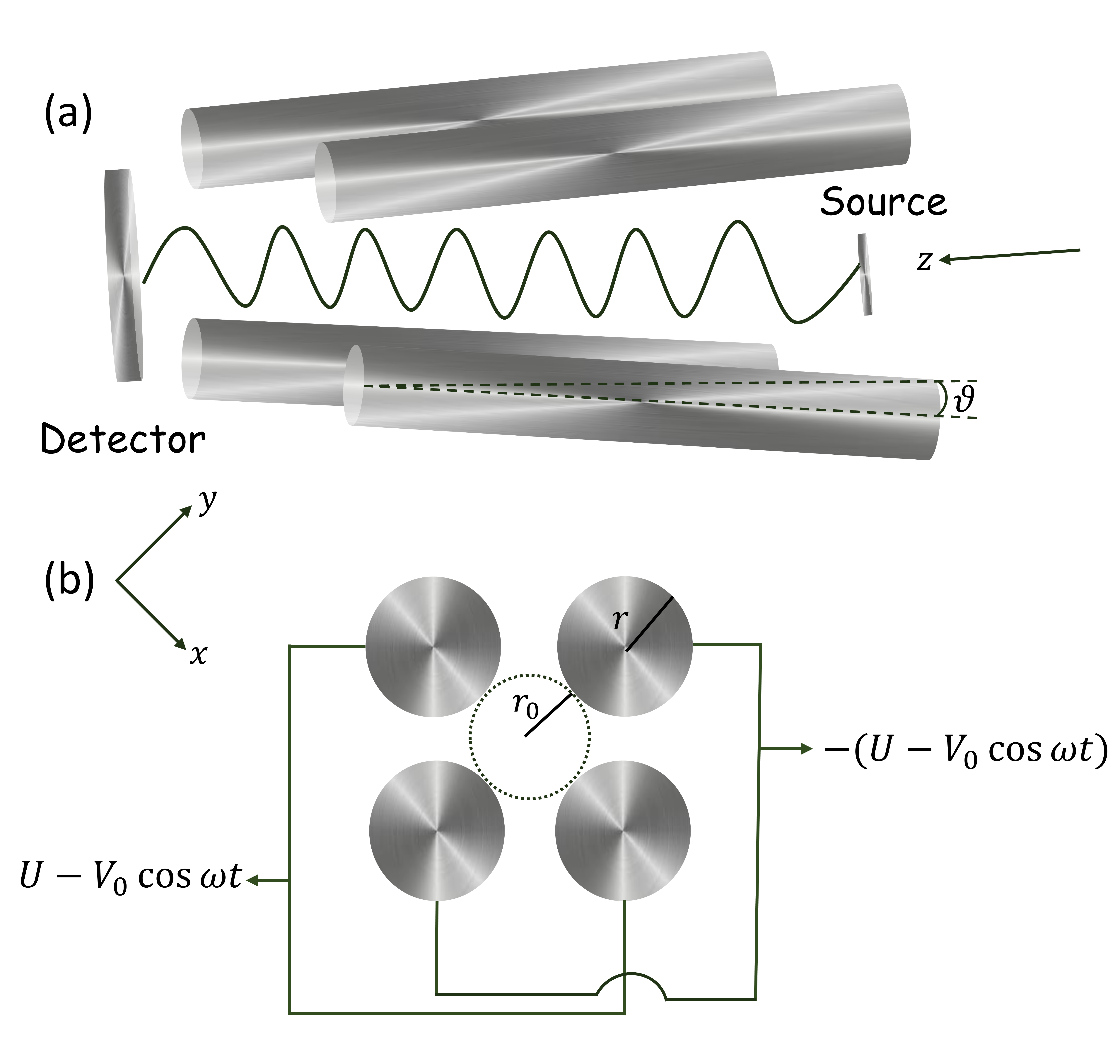}
    \caption{(a) Schematic of a quadrupole mass filter with a tapered rod configuration, in which all four rods are inclined at an angle $\vartheta$ with respect to the parallel arrangement. A flared geometry can be realized by interchanging the positions of the ion source and the detector. (b) Cross-sectional view in the $xy$ plane, illustrating the four-rod configuration along with the electrical connections at the entrance end.}
    \label{fig:1}
\end{figure}

\section{Varying Mathieu parameters}

The quadrupole potential in the radial plane of a QMF with parallel rods, for the biasing configuration schematically depicted in Fig.~\ref{fig:1}(b), is given by
\begin{equation}
\Phi(x,y,t)=\frac{\left(U-V_{0}\cos\Omega t\right)}{r_0^2}(x^2-y^2)
\end{equation}
where $r_0$ is the field radius, which remains constant for an ideal parallel rod configuration. 

The equations of motion for an ion of mass $m$ and charge $e$ along the transverse directions are governed by Mathieu equation,
\begin{equation}
\frac{d^2u}{d\tau^2}+(a_u-2q_u\cos2\tau)u=0
\end{equation}
where $u=x,y$, $\tau=\Omega t/2$ 
and the Mathieu parameters are defined as
\begin{equation}
a=\frac{8eU}{mr_0^2\Omega^2}, \quad q=\frac{4eV}{mr_0^2\Omega^2}
\end{equation}

In the ideal parallel rod configuration, $a$ and $q$ remain constant along the ion trajectory. Geometrical imperfections in the parallel rod configuration have long been a concern in fabrication, as they lead to variations in the field radius $r_0$ along the length of the QMF. Consequently, a different situation arises in which the Mathieu parameters are no longer constant but vary appreciably during the ion’s transit through the device. In tapered and flared geometries, as shown schematically in Fig.~\ref{fig:1}(a) where all four rods are uniformly inclined with respect to the parallel configuration, $r_0$ changes progressively along the axis, resulting in a corresponding modification of the quadrupole potential. Here, at the entrance rod-to-field radius ratio, $r/r_0$ ($\eta) = 1.13$ and $r_0=5$ mm.
 
For a tilting angle $\vartheta$ relative to the parallel configuration, the field radius at an axial position $z$ from the entrance of the QMF can be expressed as $r(z)=r_0+z\tan\vartheta$. This leads to a progressive decrease in the field radius for tapered geometries ($\vartheta<0$), and a corresponding increase for flared geometries ($\vartheta>0$). For a small tilting angle of $0.05\degree$, the resulting variation in field radius is $\Delta r\simeq8.73\times10^{-4}$~mm per millimeter of QMF length.

The field radius $r_0$ in eq.~1 and eq.~3 is replaced by the position-dependent radius $r(z)$, resulting in Mathieu parameters that vary along the axial direction of the QMF. An ion with an axial velocity component $v_z$ travels a distance $z=v_zt=2v_z\tau/\Omega$ after entering the QMF. Consequently, the axial variation in the field radius—and hence in the Mathieu parameters—can be equivalently expressed as a time-dependent variation.

The effective field radius can therefore be expressed in terms of $\tau$ as,
\begin{equation*}
r(\tau)=r_0\left(1+\alpha\tau\tan\vartheta\right),
\end{equation*}
where $\alpha=2v_z/\Omega r_0$. Substituting this into Eq.~(1), the modified quadrupole potential can be written as,
\begin{equation}
\Phi(x,y,t)=\frac{\left(U-V_{0}\cos2\tau\right)}{p(\tau)r_0^2}(x^2-y^2),
\end{equation}
where $p(\tau)$ is a correction factor defined as
\begin{equation}
    p(\tau)=\left(1+\alpha\tau\tan\vartheta\right)^2.
\end{equation}

The equations of motion for ions in a tilted quadrupole then become
\begin{equation}
\frac{d^2u}{d\tau^2}+
\frac{1}{p(\tau)}\left(a_u-2q_u\cos2\tau\right)u=0.
\end{equation}

The time-of-flight of an ion through the QMF of length $L$ is $t_{max}=L/v_z$. In terms of the dimensionless parameter $\tau$, the maximum value is given by
\begin{equation}
    \tau_{max}=\frac{\Omega L}{2v_z}=\frac{\Omega L}{2}\sqrt{\frac{m}{2E}},
\end{equation}
where $E$ is the longitudinal energy of the ion.

\section{First stability zone}

In an ideal quadrupole mass filter with parallel rods, ions experience a constant pseudo-potential depth throughout their time of flight. In contrast, tapered or flared geometries, characterized by convergent or divergent rod configurations, introduce an axial variation in the effective potential, resulting in a position-dependent pseudo-potential depth. Consequently, ions undergo a continuous evolution of their stability conditions as they propagate through the device, rather than operating at a fixed point in the ($a,q$) space. This leads to a significant modification of the stability region.

\begin{figure}[h]
    \centering
    \includegraphics[width=1\linewidth]{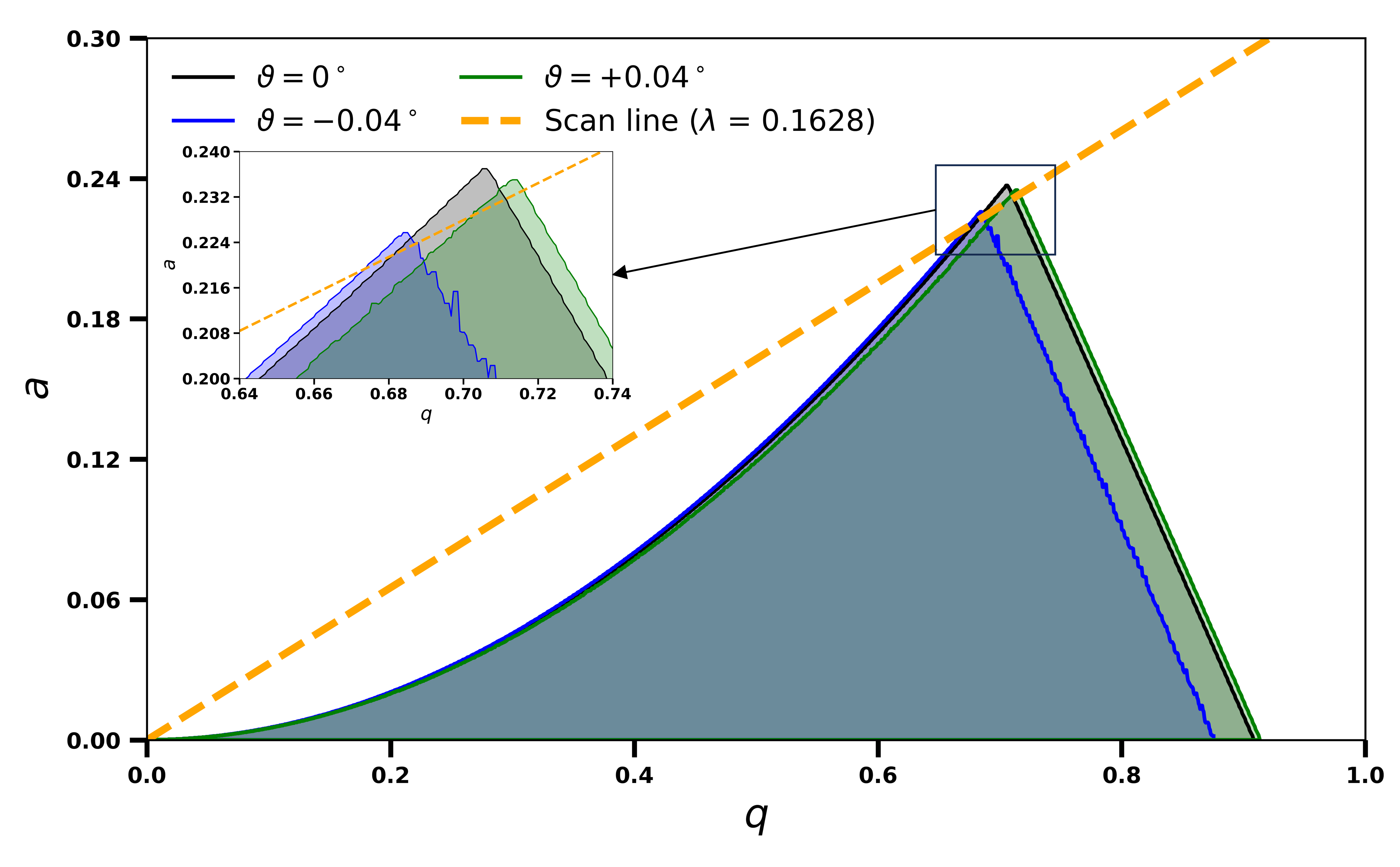}
    \caption{First stability region computed using the RK45 method for convergent and divergent rod configurations ($\vartheta=\pm0.04\degree$), in comparison with the parallel rod setup ($\vartheta=0\degree$). The inset shows an enlarged view of the apex region of the stability diagram, relevant for mass spectrometric operation.}
    \label{fig:2}
\end{figure}

To investigate the impact of rod tilting, ion trajectories are computed using a numerical integration scheme based on the fourth–fifth order Runge–Kutta (RK45) method. For each pair of Mathieu parameters ($a,q$), the equations of motion are integrated over the full time of flight through the quadrupole ($0 \le \tau \le \tau_{max}$). The simulations are performed for Ca$^{+}$ ions with a longitudinal energy of $0.5$~eV propagating through a QMF of length $160$~mm operated at a driving frequency $\Omega/2\pi = 2$~MHz. Under these conditions, $\tau_{\max} \simeq 649$, corresponding to a time of flight of approximately $103~\mu$s. This implies that the variation of the geometric correction factor $p(\tau)$, and hence of the effective Mathieu parameters, remains small over a single RF cycle (period $0.5~\mu$s). An ion is considered stable if its transverse displacement remains within the effective quadrupole aperture over the entire propagation length; otherwise, it is classified as unstable. This approach enables the construction of a modified stability diagram that captures the effects of axially varying fields arising from non-parallel rod geometries.

Fig.~\ref{fig:2} presents the simulated stability diagrams for inward and outward rod tilting of $0.04\degree$, in comparison with the ideal configuration of parallel rods. It is observed that, for the tapered geometry ($\vartheta=-0.04\degree$), the apex of the stability zone shifts toward lower values of $q$, whereas for the flared geometry ($\vartheta=0.04\degree$), it shifts toward higher values of $q$. This behavior is monotonic: increasing convergence of the rod set leads to a progressive shift of the apex toward lower $q$ values, while increasing divergence results in a corresponding shift toward higher $q$ values. This trend can be understood from the axial variation of the field radius. In the tapered geometry, the field radius decreases along the axis, resulting in an increase in the effective Mathieu parameters. Conversely, in the flared geometry, the field radius increases, leading to a corresponding decrease in the effective Mathieu parameters. In both tapered and flared configurations, the region above the constant mass scan line exhibits a noticeable contraction, indicating a potential enhancement in baseline resolution. 

\begin{figure}[h]
    \centering
    \begin{subfigure}{1.0\linewidth}
        \centering
        \includegraphics[width=\linewidth]{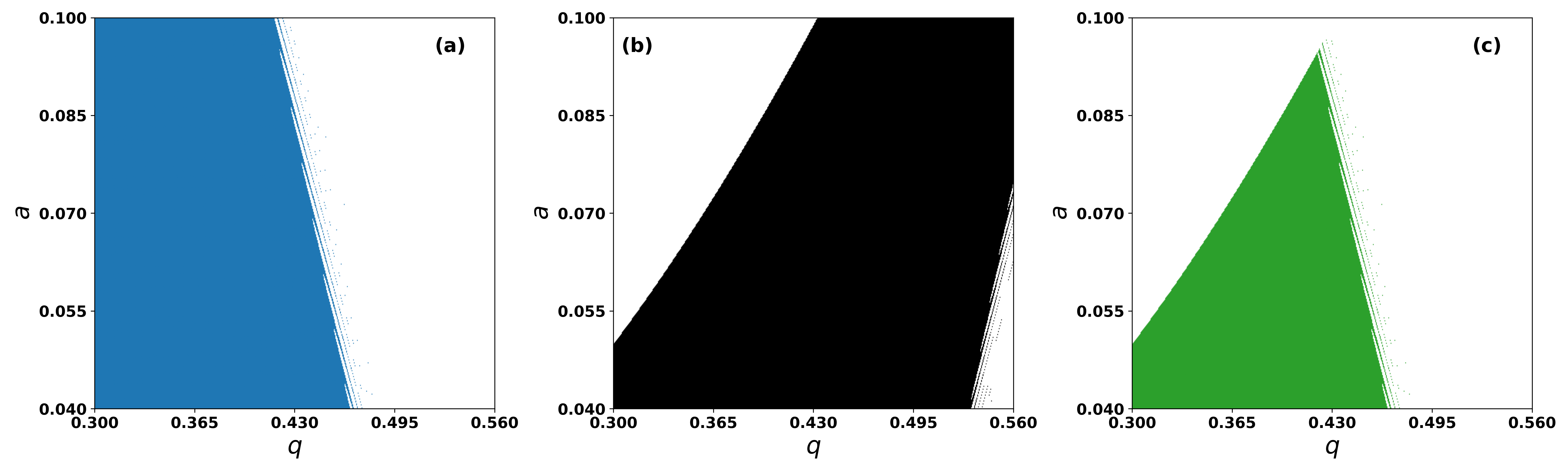}
    \end{subfigure}
    \vspace{0.5cm}
    \begin{subfigure}{1.0\linewidth}
        \centering
        \includegraphics[width=\linewidth]{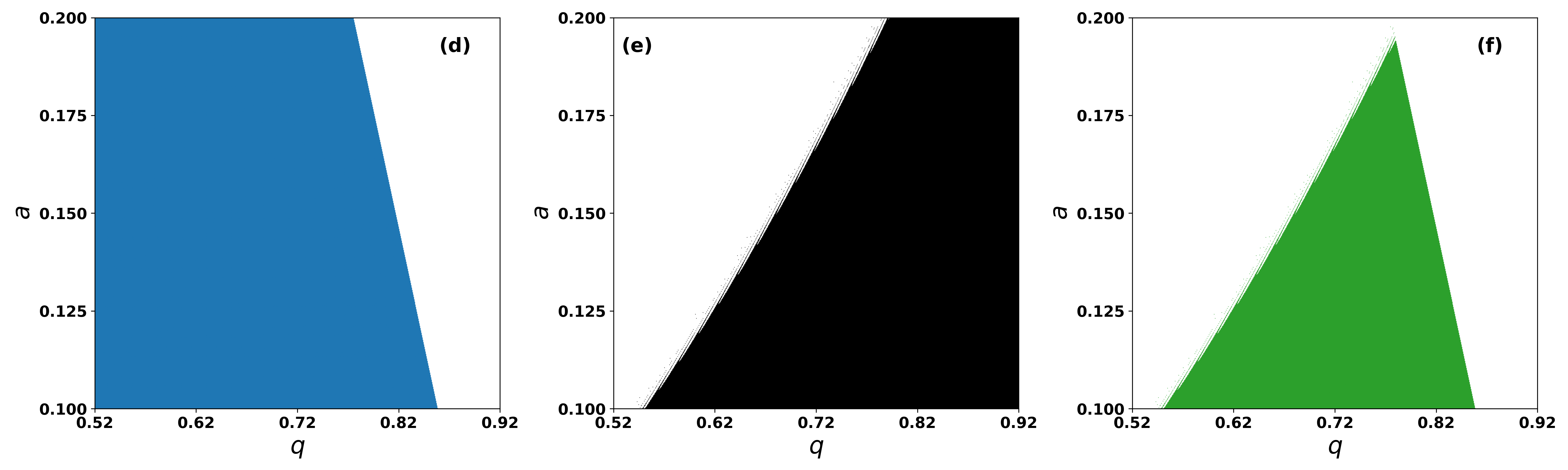}
    \end{subfigure}
    \caption{High resolution scan near the apex of the first stability zone: (a) X-motion stability, (b) Y-motion stability and (c) combined stability in the radial plane for tapered configuration ($\vartheta=-0.5\degree$). (d) X-motion stability, (e) Y-motion stability and (f) combined stability in the radial plane for flared configuration ($\vartheta=0.5\degree$).}
    \label{fig:3}
\end{figure}

A high-resolution simulation of the stability boundaries near the apex — the region of primary mass spectrometric relevance shown in Fig.~\ref{fig:3} — reveals a clear distinction in their behavior for tapered and flared geometries at a larger tilting angle of $0.5\degree$. For the tapered configuration, both the $x$- and $y$-stability boundaries exhibit noticeable contraction and become less sharply defined towards the right side of the stability region ~(Fig.~\ref{fig:3}(a), (b)). In contrast the $x$- stability boundary remains well-defined~(Fig.~\ref{fig:3}(d)), whereas the $y$- stability boundary becomes increasingly diffuse at higher ($a,q$) values~(Fig.~\ref{fig:3}(e)) in the flared geometry. This indicates a gradual transition from stable to unstable ion motion.

The axial variation in the field radius introduces a finite change in $a$ and $q$ values along the ion trajectory, causing ions to approach the stability boundary gradually. As a result, the transition between stable and unstable trajectories is no longer abrupt but spread over a range of $(a,q)$ values, leading to a diffused stability boundaries as depicted in Fig.~\ref{fig:3}. The extent of this diffusion increases with the angle $\vartheta$, as larger tilting produces stronger axial gradients in the effective field radius and hence a more pronounced variation in Mathieu parameters.

\begin{figure}
    \centering
    \includegraphics[width=0.6\linewidth]{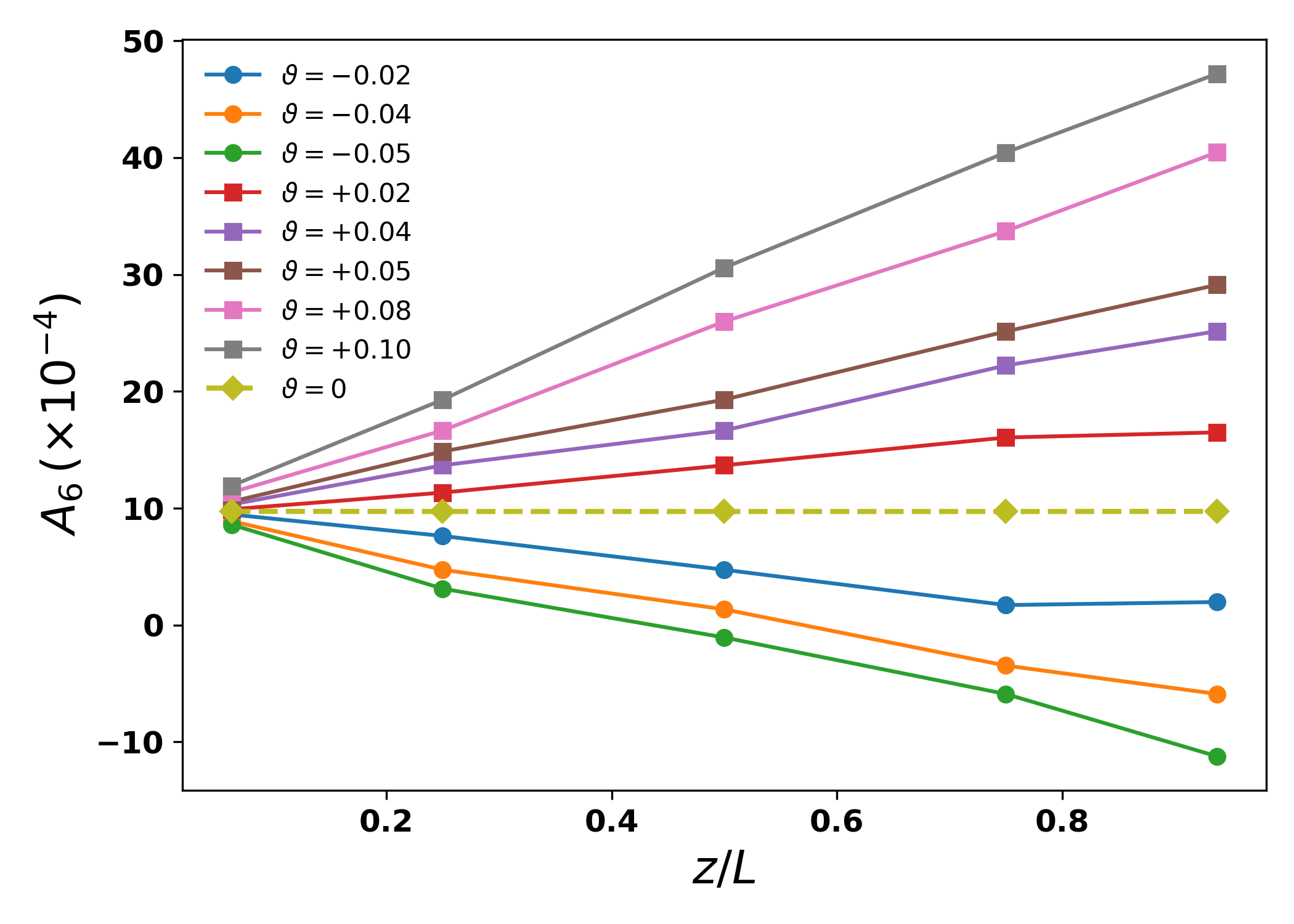}
    \caption{Variation of the higher-order multipole coefficient $A_6$ along the quadrupole length for different tapered and flared geometric configurations.}
    \label{fig:4}
\end{figure}

\section{Multipole amplitudes}

QMFs with parallel cylindrical rods inherently exhibit higher-order multipole components, most notably the dodecapole and icosapole terms, whose strengths are governed by the rod-to-field radius ratio ($\eta$). In tapered or flared geometries, this ratio varies continuously along the axial direction, leading to a corresponding axial variation in the multipole coefficients. To quantify this effect, the radial potential distribution is simulated using SIMION at different axial positions for a given geometry ($\vartheta=$ constant). The multipole coefficients are then extracted by fitting the computed potential surfaces using the methodology established in our earlier studies~\cite{DUTTA2026117621}. Fig.~\ref{fig:4} illustrates the variation of the dodecapole coefficient ($A_6$) as a function of axial position for different geometries. In contrast, the icosapole coefficient ($A_{10}$) does not exhibit significant variation along the QMF axis.

It is important to note that no octupole component arises in these tapered or flared configurations, as the rotational symmetry around the QMF axis is preserved\cite{dawson2013quadrupole}. This is further corroborated by the transmission characteristics, which remain invariant under reversal of the DC polarity.

\section{Transmission characteristics}

The transmission characteristics of a quadrupole mass filter with tapered and flared rod geometries are investigated within the first stability region using ion trajectory simulations carried out in SIMION. The tilting angle is systematically varied to control the axial rate of change of the inscribed radius. The simulations are performed using a monoenergetic ion beam consisting of $1000-5000$ ions, each with a mass-to-charge ratio of $m/z = 40$ and a longitudinal kinetic energy of $0.5~\mathrm{eV}$. The ions are injected at the entrance of the quadrupole, with their initial positions randomly distributed within a circular region of radius $0.1r_0$ about the central axis, while the initial transverse velocity components are set to zero. To average out the RF phase effects, the ion injection times are uniformly distributed over one complete RF cycle. The QMF length is $160~\mathrm{mm}$, and the system is operated at a frequency of $2~\mathrm{MHz}$.

Small tapering or flaring angles are considered to represent realistic mechanical tolerances from the ideal configuration of parallel rods. Initially, a fixed scan line ($\lambda=a/2q=0.1628$) is selected, which passes through the stability region obtained from the numerical solution of the ion equations of motion using the RK45 method for all geometries under investigation~(Fig.~\ref{fig:2}).

\begin{figure}[h]
    \centering
    \begin{subfigure}{0.48\linewidth}
        \centering
        \includegraphics[width=\linewidth, height=5cm]{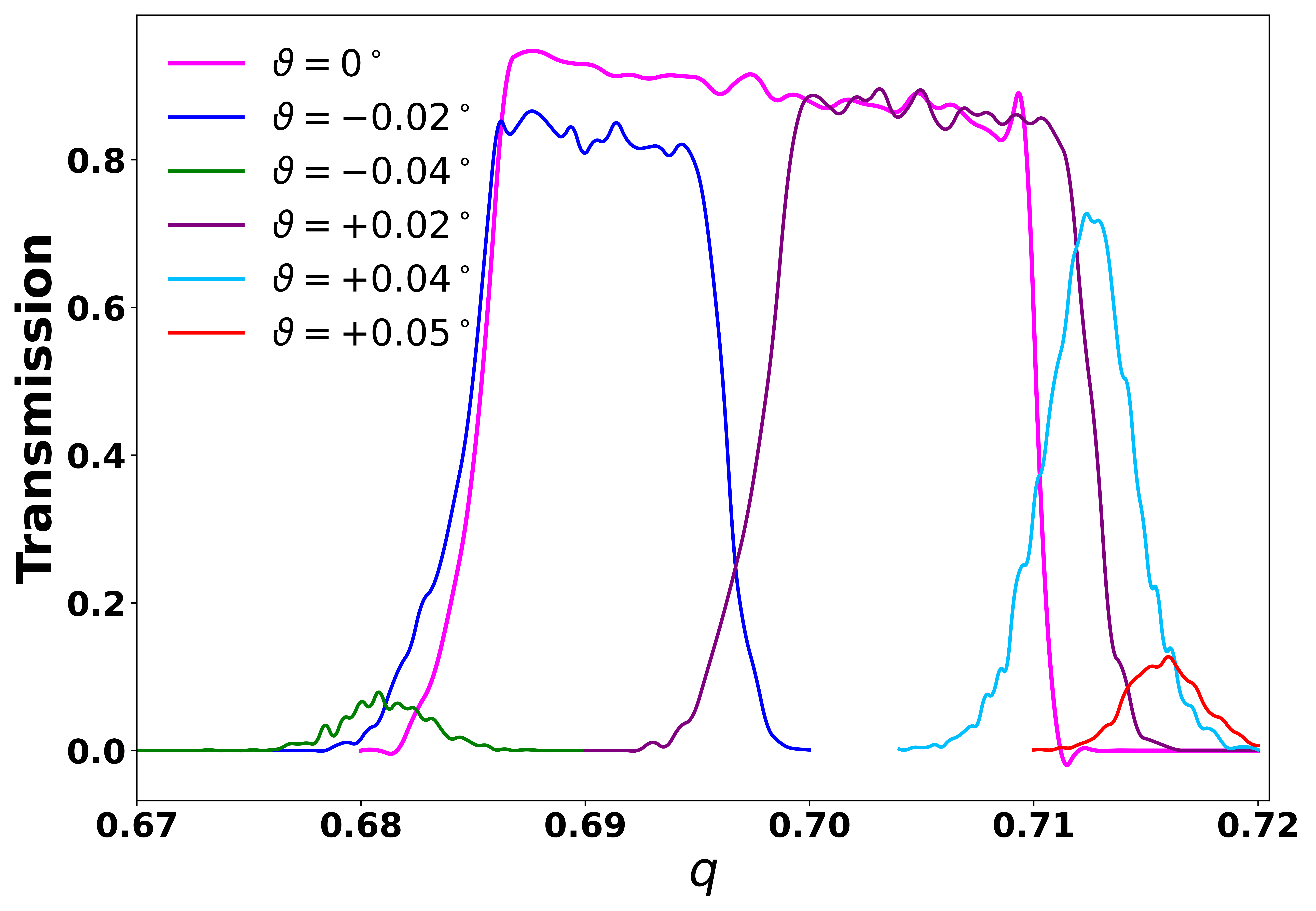}
        \caption{}
    \end{subfigure}
    \hfill
    \begin{subfigure}{0.48\linewidth}
        \centering
        \includegraphics[width=\linewidth, height=5cm]{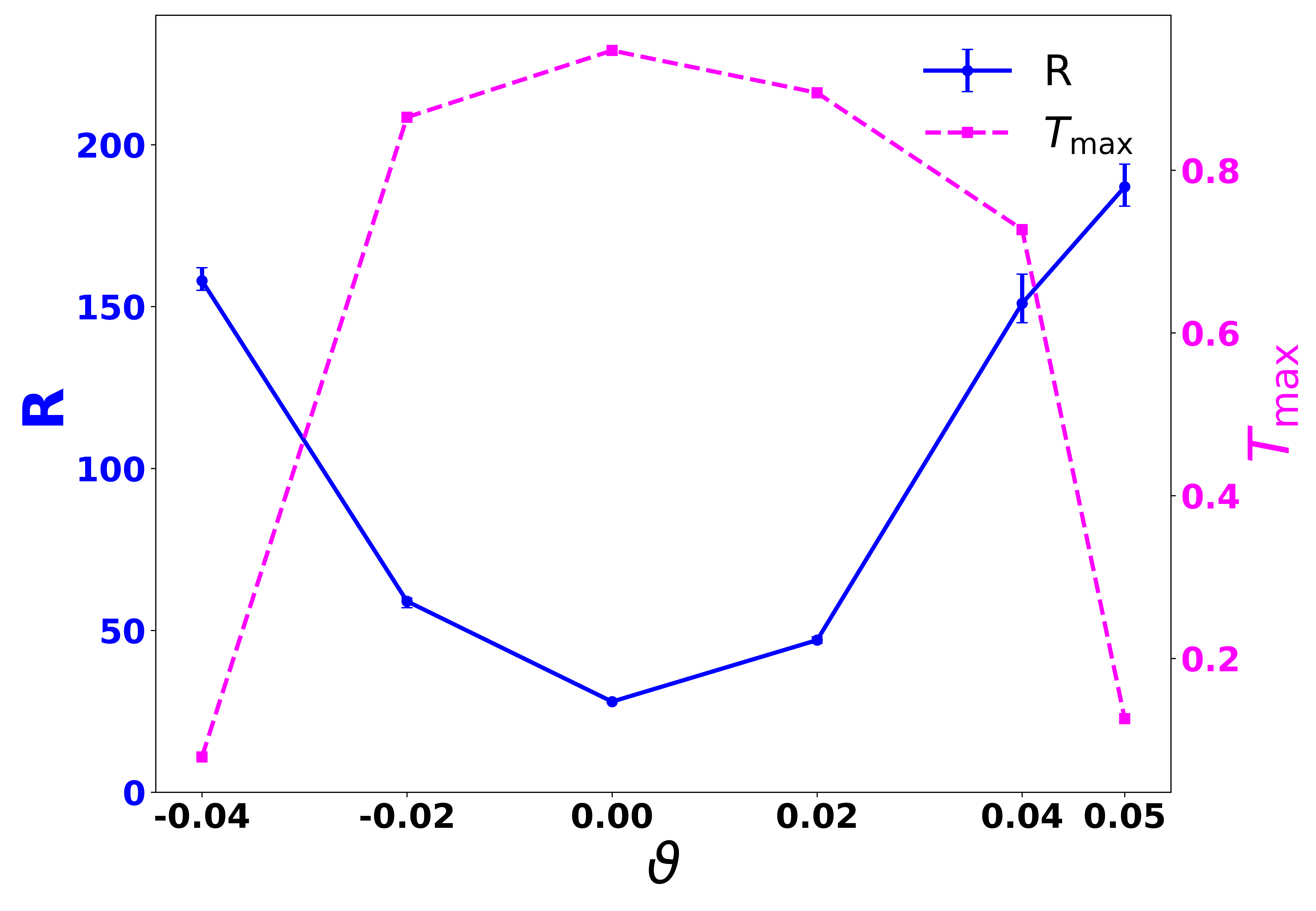}
        \caption{}
    \end{subfigure}
    \caption{(a) Transmission characteristics of a quadrupole mass filter for tapered and flared rod configurations at a fixed scan parameter ($\lambda=0.1628$) for tilting angles $\vartheta=\pm0.02\degree, \pm0.04\degree, +0.05\degree$ along with the parallel rod geometry ($\vartheta=0\degree$). (b) Resolution $R$(blue) and peak transmission $T_{\max}$(magenta) of the QMF as a function of tilt angle $\vartheta$. The error bar in $R$ is evaluated following the same approach as adopted in our earlier work~\cite{DUTTA2026117621}.}
    \label{fig:5}
\end{figure}

The transmission profiles shown in Fig.~\ref{fig:5}(a) illustrate the influence of tapered and flared geometries on QMF performance within the first stability region for a given operating condition ($\lambda=0.1628$). For the parallel rod configuration, the transmission curve exhibits a relatively broad plateau with high transmission, characteristic of ion motion well within the stability region. The transmission peak shifts to lower values of $q$ with increasing $\vartheta$ for the tapered (convergent) geometry, whereas it shifts to higher values of $q$ with increasing $\vartheta$ for the flared (divergent) geometry. The observed shift in the transmission peak with increasing tilt angle ($\vartheta$) is consistent with the corresponding modifications in the stability diagram~(Fig.~\ref{fig:2}).

As shown in Fig.~\ref{fig:5}(a), the transmission decreases slightly at $\vartheta=-0.02\degree$ and drops sharply at $\vartheta=-0.04\degree$; beyond this, it becomes negligibly small for the given operating condition in the tapered configuration. In contrast, for the flared configuration, the reduction in transmission is more gradual up to $\vartheta=0.04\degree$ followed by a sharp decrease at $\vartheta=0.05\degree$, as also evident from Fig.~\ref{fig:5}(a). For flaring angles beyond $\vartheta=0.05\degree$, the transmission becomes negligible under the same operating conditions.

The change in the peak structure observed in the tilted configurations indicates a corresponding variation in the resolution with the angle of tilting. For a quantitative assessment, the resolution is defined as $R=q_0/\Delta q$ and is evaluated from the transmission curves for each value of $\vartheta$ at $\lambda=0.1628$. Here $q_0$ denotes the peak position (or the midpoint of the transmission profile, where applicable), and $\Delta q$ represents the full width at half maximum (FWHM) of the transmission contours. The variation in $R$ and peak transmission ($T_{max}$) with $\vartheta$ are shown in Fig.~\ref{fig:5}(b). As evident from Fig.~\ref{fig:5}(b), the resolution increases for both tapered and flared configurations, with comparable levels of enhancement. However, the flared geometry maintains significantly higher transmission for a similar gain in resolution, indicating a more favorable balance between transmission and resolution under the given operating condition.

The trade-off between transmission and resolution, shown in Fig.~\ref{fig:5}(b), can be qualitatively understood from the evolution of the stability diagram in tapered and flared geometries~(Fig.~\ref{fig:2}). The chosen mass scan line ($\lambda=0.1628$) intersects a narrower region near the apex of the stability boundary for both configurations, as evident from Fig.~\ref{fig:2}, thereby indicating an enhancement in baseline resolution.

The decrease in the effective field radius along the axial direction in the tapered geometry leads to a progressively stronger radial electric field toward the exit of the QMF. While this enhanced field improves the radial confinement, it also imposes more stringent stability conditions near the exit and and strengthens the coupling between axial and radial motion. Consequently, ions with marginal stability are more likely to be ejected from stable trajectories, resulting in a reduction in overall transmission, as observed in Fig.~\ref{fig:5}. At the same time, the increased field gradient sharpens the transmission profile, yielding a reduced FWHM ($\Delta q$) and, therefore, an improvement in resolution.

In contrast, for the flared geometry, the effective field radius increases along the axial direction, resulting in a gradual weakening of the electric field toward the exit. This reduction in field strength relaxes the stability constraints, allowing a larger fraction of ions to remain on stable trajectories throughout their propagation. 
The gradual axial variation of the radial field modifies the phase-space evolution of the ions, leading to an improvement in resolution without a significant compromise in transmission for smaller angle of tilting.

Overall, the results demonstrate that tapered geometry enhances resolution at the expense of transmission due to increased ion loss near the exit, whereas flared geometry provides a more balanced behavior, maintaining high transmission while achieving moderate resolution improvement. These observations highlight the critical role of axial field variation in controlling the transmission–resolution trade-off in QMFs with tilted geometry.

For a consistent comparison of QMF performance in the presence of mechanical deviations, transmission studies are often conducted at a fixed baseline resolution~\cite{DUTTA2026117621}. However, the definition of a constant baseline resolution is subject to inherent uncertainty, particularly due to the diffuse nature of the stability boundaries, which are typically derived under the assumption of an ideal quadrupole field. In such calculations, the stability diagram is constructed by assuming that the longitudinal energy of the ions remains constant throughout their trajectory. This assumption is not strictly valid, as coupling between radial and axial motion can lead to energy exchange between these degrees of freedom, an effect that has been exploited in studies of single-ion heat engines in tapered ion trap geometries\cite{Abah2012,Rossnagel2016}.

Furthermore, ion trajectory simulations using SIMION reveal a distribution in the time of flight of ions traversing the QMF, whereas a constant time of flight is implicitly assumed in the construction of stability diagrams. This discrepancy highlights a limitation of stability-diagram-based analyses when applied to systems with axially varying fields. Consequently, comparing the QMF performance solely on the basis of a constant baseline resolution may not provide a fully realistic description. In the following, a more representative criterion—based on constant peak transmission—is adopted to assess the performance of QMFs with tapered and flared geometries.

\begin{figure}
    \centering
    \includegraphics[width=1\linewidth]{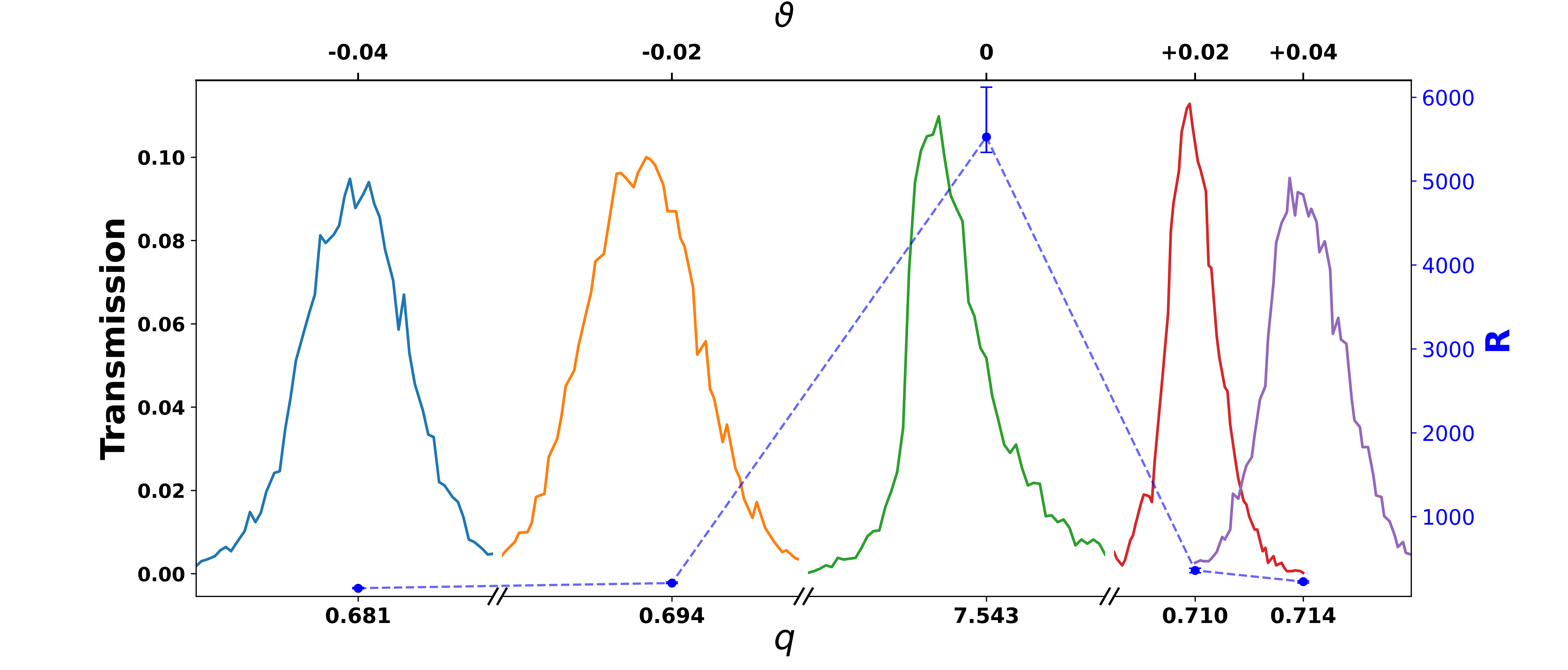}
    \caption{Transmission characteristics of tapered and flared QMFs ($\vartheta=0\degree, \pm 0.02\degree, \pm 0.04\degree$). The dashed blue lines tracks the variation in $R$ with $\vartheta$. The error bar in $R$ is evaluated following the same approach as adopted in our earlier work~\cite{DUTTA2026117621}.}
    \label{fig:6}
\end{figure}

Transmission studies are performed for different tilting angles at a fixed peak transmission of approximately $10\%$, with the scan-line slope ($\lambda$) adjusted accordingly. The resulting characteristics are shown in Fig.~\ref{fig:6}, along with the corresponding variation in resolution as a function of the tilting angle. As evident from Fig.~\ref{fig:6}, the parallel-rod configuration produces a sharp transmission peak and yields the highest resolution for the specified peak transmission. In contrast, both tapered ($\vartheta<0\degree$) and flared ($\vartheta>0\degree$) geometries exhibit broadened transmission peaks, accompanied by a monotonic decrease in resolution with increasing $|\vartheta|$. Thus, for a fixed transmission level, any deviation from the parallel rod configuration leads to a degradation in QMF performance.

\section{Conclusion}
A detailed simulation study of the performance of a quadrupole mass filter (QMF) with tapered and flared rod geometries has been presented. The progressive axial variation in the field radius is found to significantly influence ion stability during transport through the QMF. Stability diagrams, computed using the RK45 method for the ideal quadrupole field, reveal a shift in the stability apex and contraction of the stable region for tapered geometries and flared geometries. High-resolution simulations further indicate the emergence of diffuse stability boundaries, suggesting a gradual transition from stable to unstable ion motion.

Transmission simulations performed at a fixed scan line reveal a clear interplay between transmission and resolution, arising from the underlying modifications in the stability region. In tapered geometries, increasing convergence rapidly suppresses transmission, although a slight degree of tapering can yield a measurable improvement in resolution without appreciable loss in transmission. In contrast, flared geometries offer a more favorable operating regime: small tilting angles lead to a discernible enhancement in resolution while preserving high transmission. This highlights the relative advantage of flared configurations in achieving improved performance without significant compromise.
 
A realistic comparison, based on constant peak transmission, shows that even small deviations from ideal parallel rod geometry lead to a noticeable degradation in resolution.
While the present study focuses on the effects of mechanical imperfections in QMFs with parallel circular rods, the results offer broader insights for the design and operation of quadrupole-based devices. In particular, these results are relevant for ion trap configurations employing non-parallel electrode geometries, including recent proposals involving tapered or funnel-shaped ion traps for controlled ion dynamics and thermodynamic applications~\cite{Abah2012,Rossnagel2016,torrontegui2018transient,Dawkins2018,Levy2020}. Furthermore, the observed modifications in the stability characteristics point to the potential for operating QMFs in higher stability regions without the need for an applied DC component, opening avenues for alternative modes of operation.

\begin{acknowledgement}
ND thanks SERB/ANRF India (CRG/2023/001529) and BRNS India (58/14/21/2023 - BRNS12329) for funding. AD acknowledges DST Inspire for research fellowship.
\end{acknowledgement}

\section{Conflicts of Interest}
The authors declare no conflicts of interest for this manuscript.

\bibliography{reference}

\end{document}